\documentclass{appolb}
\usepackage{epsfig}
\usepackage{graphicx}% Include figure files
\usepackage{bm}% bold math
\usepackage{amsmath,epsfig}
\usepackage{natbib}
\usepackage{dcolumn}
\usepackage{graphicx}
\usepackage{amsmath}
\usepackage[hang,small,bf,centerlast]{caption} 
\begin{document}
\title{Shrinkage and spectral filtering of correlation matrices: \\ a comparison via the Kullback-Leibler distance
\thanks{Presented at  the Workshop ``Random Matrix Theory: From Fundamental Physics To Application", Krakow, Poland, May 3Ð-5, 2007.}%
}
\author{M. Tumminello$^{1}$ , F. Lillo$^{1,2,3}$, and R.N. Mantegna$^{1,3}$
%\address{$^{\times}$ INFM-CNR, Unit\`a di Palermo, Palermo, Italy}
\address{$^1$ Dipartimento di Fisica e Tecnologie Relative, Universit\`a degli Studi di Palermo, Viale delle Scienze, Edificio 18, I-90128, Palermo, Italy}
\address{$^2$ Santa Fe Institute, 1399 Hyde Park Road, Santa Fe, NM 87501, USA}
\address{$^3$ CNR-INFM, Unit\`a Operativa di Roma, Centro di Ricerca e Sviluppo SOFT} 
}
\maketitle
\begin{abstract}
The problem of filtering information from large correlation matrices is of great importance in many applications. We have recently proposed the use of the Kullback-Leibler distance to measure the performance of filtering algorithms in recovering the underlying  correlation matrix when the variables are described by a multivariate Gaussian distribution.
Here we use the Kullback-Leibler distance to investigate the performance of filtering methods based on Random Matrix Theory and on the shrinkage technique. We also present some results on the application of the Kullback-Leibler distance to multivariate data which are non Gaussian distributed.
\end{abstract}
  
%%%%%%%%%%%%%%%%%%%%%%%%%%%%%%%%%%%%%%%%%%%%%%%%%%%%%%%%%%%%%%%%%%%%%%%%%%%%%%%%%%%%%%%%%%%%%%%%%%%%%%%%%%
%%%%%%%%%%%%%%%%%%%%%%%%%%%%%%%%%%%%%%%%%%%%%%%%%%%%%%%%%%%%%%%%%%%%%%%%%%%%%%%%%%%%%%%%%%%%%%%%%%%%%%%%%%
\section{Introduction}
In many applications the monitoring of the dynamics of the system provides multivariate time series and often the number of monitored variables  is very high. Examples include gene expression level measurement in microarray experiments \cite{brown}, fMRI experiments \cite{Eguiluz}, analysis of economic or financial data such as firm growth rates or stock price returns \cite{Laloux1999,Plerou1999,Mantegna1999}. A common way to investigate the interaction between the variables of the system is through the cross correlation matrix. As any statistical estimator, the sample correlation matrix is  unavoidably affected by the statistical uncertainty due to the finite size of the sample. This problem becomes extremely important when the number of investigated variables is comparable with the number of records of each variable. To cope with the problem of the statistical uncertainty of the sample correlation matrix one needs to introduce filtering methods able to remove from the correlation matrix at least part of the noise.  Many techniques have been proposed in the literature in order to filter out information from the correlation matrix.  However, unless one knows in advance the model describing the system dynamics, it is difficult to asses the goodness of the filtering procedures. Recently \cite{Tumminello2007} we have proposed the use of the Kullback-Leibler (KL) distance as a method of assessing the performance of correlation matrix filtering procedures. There are several reasons why we believe KL distance is a good performance estimator. The main reason is that we proved \cite{Tumminello2007} that for Gaussian distributed variables the expected values of the KL distance are independent from the underlying model. This fact allowed us to devise a method to asses the performance of the filtering method in recovering the underlying model without having any knowledge on the model itself. 

In this paper we consider filtering procedures based on Random Matrix Theory (RMT), hierarchical clustering and shrinkage and we use the KL distance to evaluate their  performance. We consider both artificial and real data samples. Finally we present an extension of the KL distance to an important class on non-Gaussian distribution, specifically the multivariate Student's t-distribution.

\section{Kullback-Leibler distance for Gaussian variables}
The KL distance (see for instance \cite{Kullback,Cover}) or mutual entropy is a measure of the distance between two probability densities, say $p$ and $q$, which is defined as $K(p,q)=E_p\left[\log{\left( p/q \right)}\right]$,
where $E_p[\, . \, ]$ indicates the expectation value with respect to the probability density $p$. The KL distance is asymmetric since the expectation value is evaluated according to the distribution $p$. 

Here we consider the KL distance between multivariate probability distributions and we indicate with $n$ the dimension of the space spanned by the variables. 
Let us consider first the case of multivariate Gaussian variables. Without loss of generality we assume that the variables have zero mean and unit variance. In this case, the Gaussian multivariate probability density function $P({\bf \Sigma},X)$  is completely defined by the correlation matrix ${\bf \Sigma}$ of the system.
Given two different probability density functions $P({\bf \Sigma}_1,X)$ and $P({\bf \Sigma}_2,X)$, we have
%
 %
%\begin{widetext}
\begin{eqnarray}
\label{kullbackMulti}
K(P({\bf \Sigma}_1,X),P({\bf \Sigma}_2,X))
=\int{P({\bf \Sigma}_1,X) \log\left[ \frac{P({\bf \Sigma}_1,X)}{P({\bf \Sigma}_2,X)}\right] dX}=\nonumber \\
=\frac{1}{2} \left[\log{\left(\frac{|{{\bf \Sigma}_2}|}{|{{\bf \Sigma}_1}|}\right)}+\text{tr}\left({{\bf \Sigma}_2^{-1} {\bf \Sigma}_1 }\right)-n\right],
\label{kullbackGaussian}
\end{eqnarray}
where ${|{{\bf \Sigma}}|}$ indicates the determinant of ${\bf \Sigma}$.
From now on we indicate \\$K(P({\bf \Sigma}_1,X),P({\bf \Sigma}_2,X))$ simply with $K({\bf \Sigma}_1,{\bf \Sigma}_2)$. 

Consider the Pearson sample correlation matrix ${\bf C}$ obtained from the observation of the $n$ variables  each for  $T$ records. The sample correlation matrix is different from the true correlation matrix of the system. 
The Pearson estimator of the correlation matrix has the advantage that sample covariance matrices of finite variance variables belong to the ensemble of Wishart random matrices and many statistical properties of Wishart matrices are known \cite{Mardia}. Since different realizations of the process give rise to different sample correlation matrices, a KL distance having one or two sample correlation matrices as arguments is a function of one or two random matrices.  We investigated  the statistical properties of KL distance involving sample correlation matrices of multivariate Gaussian random variables in Ref. \cite{Tumminello2007}. 
  
Let ${\bf C}_1$ and ${\bf C}_2$ be two sample correlation matrices obtained from two independent realizations of the system both of length $T$.
By making use of the theory of Wishart matrices \cite{Mardia} we obtain \cite{Tumminello2007} that
\begin{eqnarray}
\label{kullexpecSigS1}
&&E\left[K({\bf \Sigma},{\bf C}_1)\right]=\frac{1}{2} \left \{n\log{\left(\frac{2}{T}\right)}+\sum_{\ell=T-n+1}^{T}{\left[\frac{\Gamma^{\prime}(\ell/2)}{\Gamma(\ell/2)}\right]}+\frac{n (n+1)}{T-n-1}\right\},\\
\label{kullexpecS1Sig}
&&E\left[K({\bf C}_1,{\bf \Sigma})\right]=\frac{1}{2} \left \{n\log{\left(\frac{T}{2}\right)}-\sum_{\ell=T-n+1}^{T}{\left[\frac{\Gamma^{\prime}(\ell/2)}{\Gamma(\ell/2)}\right]}\right\}\\
\label{kullexpecS1S2}
&&E\left[K({\bf C}_1,{\bf C}_2)\right]=\frac{1}{2} \frac{n (n+1)}{T-n-1},
\end{eqnarray}
where 
$\Gamma(x)$ is the usual Gamma function and $\Gamma^{\prime}(x)$ is the derivative of $\Gamma(x)$. 
We also obtained \cite{Tumminello2007} the asymptotic expectation value of the standard deviation of $K({\bf C}_1,{\bf \Sigma})$ by using the Bartlett statistics \cite{Bartlett54}. Specifically if $T\gg1$, $n\gg1$ and $Q=T/n\gg1$ we infer that the standard deviation of $K({\bf C}_1,{\bf \Sigma})$ is $\sigma_K\simeq1/(2Q)$.\\
 
The most important property of the expectation values given in Eq.s (\ref{kullexpecSigS1}-\ref{kullexpecS1S2}) is that they are independent of ${\bf \Sigma}$, i.e. they are independent of the specific true correlation matrix. This fact implies that (i) the KL distance is a good measure of the statistical uncertainty of correlation matrix which is due to the finite length of data series and (ii) the expected value of the KL distance is known also when the underlying model hypothesized to describe the system is unknown. 
%It is worth noting that other matrix distances, such as the Frobenius distance, do not have the model independence property. However there are distances, such as $\log(|{\bf \Sigma}_2|/|{\bf \Sigma}_1|)$ which have expectation values independent from the model.

\section{Comparison of filtering procedures}

The KL distance can be used to quantify and compare the performance of different filtering procedures of correlation matrices \cite{Tumminello2007}. A good filtering procedure should have two important properties: (i) being able to remove the ``right" amount of noise from the data in order to recover the signal and (ii) produce filtered matrices which are stable when one makes different observations of the same system. These two requirements are often in competition one with the other.  
In real cases one does not know the true correlation matrix, therefore it seems impossible to know whether a filtering procedure is removing the right amount of noise. However the above mentioned property of the expected value of the KL distance of being independent from the model correlation matrix can be used to estimate the goodness of the filtering procedure. The proposed procedure to evaluate the performance of a filtering procedure is the following.

Suppose we are given with a data sample ${\bf X}$ and we have our favorite filtering procedure. We propose to generate $M$ bootstrap replicas ${\bf X}_i$ ($i=1,..,M$) of the data. We then compute the sample correlation matrix ${\bf C}_i$ and apply the filtering procedure obtaining the filtered matrix ${\bf C}^{filt}_i$ to each replica ${\bf X}_i$. In order to measure the stability of the filtering procedure, we consider the average of over the replicas of the quantity $K({\bf C}^{filt}_i,{\bf C}^{filt}_j)$. An optimal filtering procedure should be perfectly stable (i.e. $\langle K({\bf C}^{filt}_i,{\bf C}^{filt}_j) \rangle=0$) because from each realization the filtering recovers the model matrix. 
In order to measure the filtered information we consider the average of $K({\bf C}_i,{\bf C}^{filt}_i)$ over the replicas. This quantity measures the information present in the sample correlation matrix ${\bf C}_i$ that has been discarded by the filtering procedure. 
We have seen above that for Gaussian variables the KL distance  $\langle K({\bf C}_i,{\bf \Sigma}) \rangle$ is different from zero and independent from the model ${\bf \Sigma}$ (see Eq.~\ref{kullexpecS1Sig}). Therefore if our filtering procedure is recovering the true underlying model we should expect that $K({\bf C}_i,{\bf C}^{filt}_i)$ is equal to the right hand side of Eq. \ref{kullexpecS1Sig}. We have thus an optimal value for both the stability and the information expected from an optimal filtering and these values are independent from the underlying model. 
We will represent the result of the analysis with a plane where the $x$ axis is related to the stability $\langle K({\bf C}^{filt}_i,{\bf C}^{filt}_j) \rangle$ and the $y$ axis is related to the information $\langle K({\bf C}_i,{\bf C}^{filt}_i)\rangle$.
In this plane the optimal point, labeled ${\bf \Sigma}$, has coordinate $x=0$ and $y$ equal to the right hand side of Eq. \ref{kullexpecS1Sig}. A filtering procedure will be considered good if the corresponding point in the stability-information plane is close to ${\bf \Sigma}$. 

There are many different filtering procedures. A widespread procedure is based on random matrix theory \cite{Metha90}. If the $n$ variables are independent and with finite variance then in the limit $T,n \to \infty$, with a fixed ratio $Q= \geq 1$, the eigenvalues of the Pearson sample correlation matrix ${\bf C}$ is bounded from above by the value $ \lambda_{max}=\sigma^2 (1+1/Q+2\sqrt{1/Q})$ where $\sigma^2=1$ for correlation matrices. In some practical cases, such as for example in finance, one finds that the largest eigenvalue $\lambda_1$ of the empirical correlation matrix is definitely inconsistent with RMT. In these cases, the null hypothesis is modified so that correlations can be explained in terms of a one factor model and $\sigma^2=1-\lambda_1/n$ \cite{Laloux1999}.
The  filtering procedure considered here works as follows  \cite{Potters2005}. One diagonalizes the correlation matrix and replaces the all eigenvalues smaller than   $\lambda_{max}$ in the diagonal matrix with their average value. Then one retransforms the modified diagonal matrix in the standard basis obtaining a matrix ${\bf H}_{RMT}$ of elements $h_{ij}^{RMT}$.
Finally, the filtered correlation matrix ${\bf C}^{RMT}$ is the matrix of elements $c_{ij}^{RMT}=h_{ij}^{RMT}/\sqrt{h_{ii}^{RMT}\,h_{jj}^{RMT}}$.

In this paper we also consider hierarchical clustering based filtering procedures  \cite{Anderberg}.  Hierarchical clustering methods allow to hierarchically organize the elements in a rooted tree or dendrogram. The whole information about the rooted tree can be stored in a $n \times n$ matrix that can be considered as the output of the filtering procedure  \cite{Anderberg}. In a recent paper we have shown that this filtered matrix is a proper correlation matrix at least when all of its elements are non negative numbers \cite{epl}. Here we consider two widespread hierarchical clustering  techniques, specifically the Single Linkage Cluster Analysis (SLCA) and the Average Linkage Cluster Analysis (ALCA)  \cite{Anderberg}. For more details about these techniques see Refs \cite{Tumminello2007,epl}.
Finally, we also consider a shrinkage filtering procedure \cite{haff,Ledoit03} in which we construct a filtered matrix as
\begin{equation}
{\bf C}^{SHR}(\alpha)=\alpha {\bf T}+(1-\alpha){\bf C},
\label{shrinkage}
\end{equation}
where $0\le\alpha\le1$ and ${\bf T}$ is a target matrix. As commonly done in financial literature, we choose the target matrix as a matrix with $t_{ii}=1$ and $t_{ij}=\langle c_{ij}\rangle$ for $i\ne j$. We estimate the performance of the shrinkage procedure for different values of $\alpha$. It is also interesting to note that there exist analytical methods to obtain the optimal value $\alpha^*$ according to a cost function  based on standard quadratic (or Frobenius) norm \cite{Schafer}.  In the figures we also show the point (labeled ${\bf C}^{SHR}(\alpha^*)$) corresponding to the value $\alpha^*$.

\begin{center}
\begin{figure}[ptb]

\begin{center}
              \hbox{
              \includegraphics[scale=0.22]{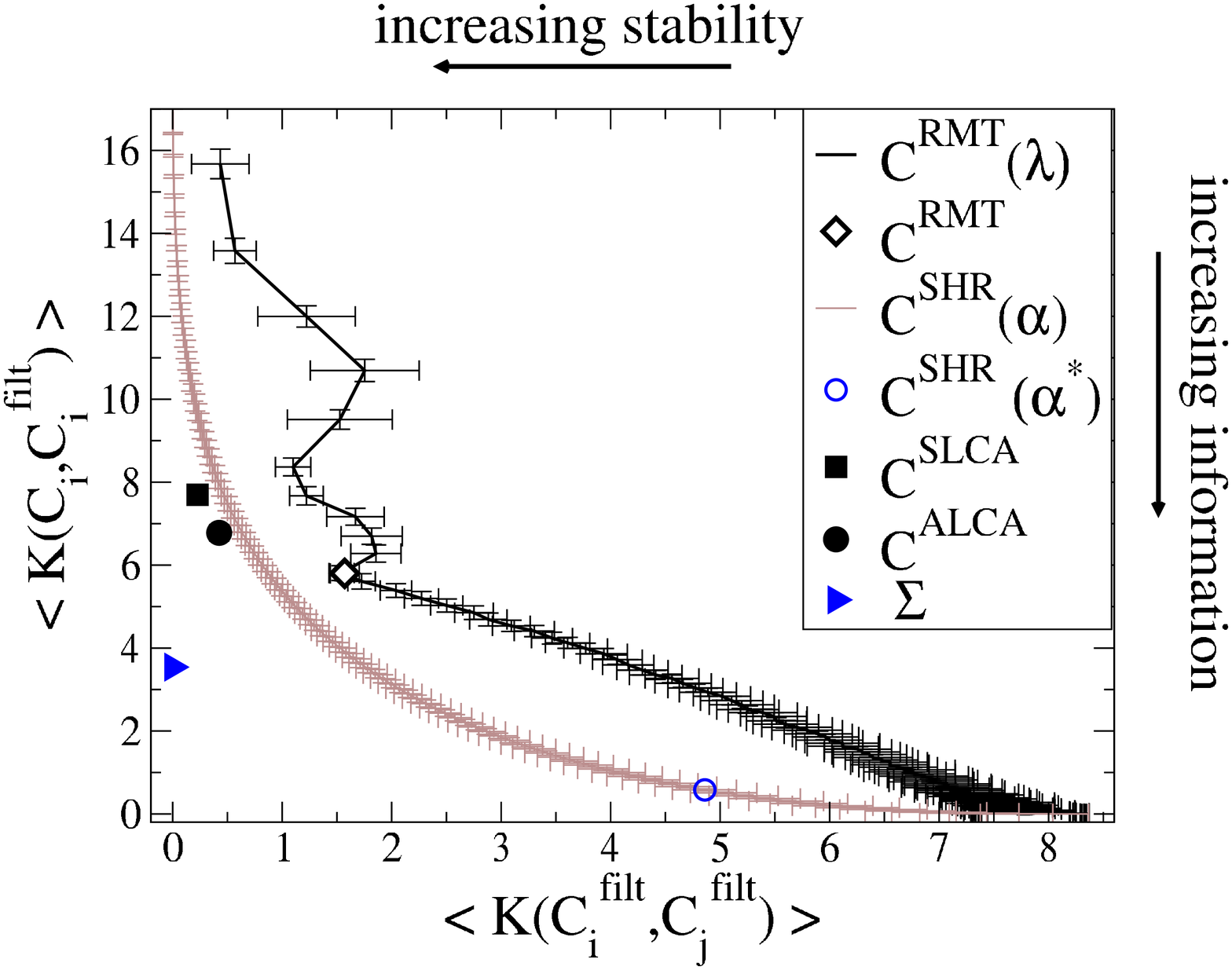}
              \hspace{. cm}
              \includegraphics[scale=0.22]{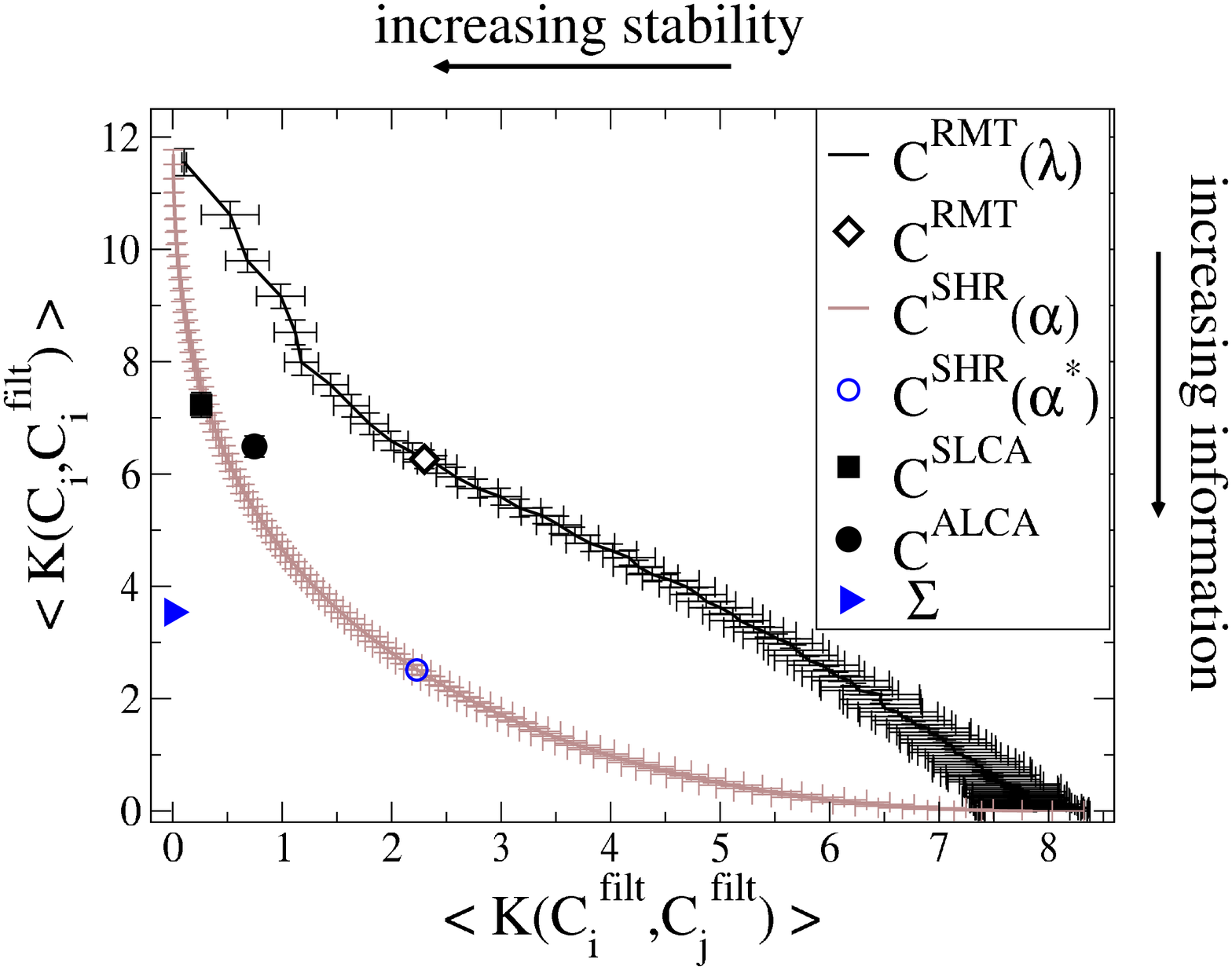}
              }
\end{center}              \caption{Stability of the filtered matrix ($x$ axis) against the amount of information about the correlation matrix that is retained in the filtered matrix ($y$ axis). 
The points labeled with ${\bf C}^{RMT}(\lambda)$ correspond to the filtering procedure keeping a fixed number of eigenvalues. The number of kept eigenvalues increases when one goes from the top left to the bottom right corner. The points labeled with ${\bf C}^{SHR}(\alpha)$ correspond to the shrinkage procedure (see Eq.~\ref{shrinkage}) and the parameter $\alpha$ goes from $0$ to $1$ when one goes from the bottom right to the top left corner.
 Left panel shows the result for a block diagonal model of $n=100$ elements divided in $12$ groups and simulated for $T=748$ points. Right panel shows the result for a hierarchically nested model of 100 elements following the HNFM with 23 factors of Ref. \cite{epl}.} 
              \label{RMT}

\end{figure}
\end{center}

In fig.~\ref{RMT} we show the KL distance in the plane stability-information for these filtering procedures applied to artificial data generated according to two different models. The left panel shows the result for a block diagonal model with $12$ blocks, whereas the right panel shows the result for a hierarchical model.  This is a Hierarchically Nested Factor Model (HNFM) with $23$ factors and it has been introduced in Ref.  \cite{epl}. In both panels we show the points corresponding to the RMT, SLCA, and ALCA filtering procedures. We also show the points corresponding to filtering procedures in which an {\it a priori}  fixed number of eigenvalues is retained and the remaining ones are set equal to their average. We also show the points corresponding to the shrinkage filtering procedure of Eq.~\ref{shrinkage} for different values of $\alpha$.
As expected when one includes more and more eigenvalues in the filtering procedure the amount of discarded information decreases and the filtered matrix becomes less and less stable. Interestingly in the block diagonal model a clear kink is observed close to the point corresponding to the filtering procedure where $12$ eigenvalues are included. The point corresponding to the kink is also the closest to the optimal point  and close to the point corresponding to the RMT filtering procedure outlined above. This result shows that for simple block diagonal models RMT and KL procedures gives consistent results.
For the hierarchical model we observe no kink when one varies the number of eigenvalues retained by the filtering procedure. This fact indicates that filtering procedures based on spectral analysis may have problems in filtering correlation matrices with a complex structure. Moreover, the number of eigenvalues retained by the RMT filtering procedure is not equal to the number of factors of the HNFM. In the case of the hierarchical model  the structure of eigenvalues and eigenvectors is definitely more complicated than the one observed for a block diagonal model. Such a structure is better recovered in the filtering by hierarchical clustering techniques according to the right panel of fig.~\ref{RMT}. Finally, the shrinkage method is capable to achieve a very good compromise between stability and information. From this analysis it is possible to extract an optimal value of $\alpha$ minimizing the distance from the point labeled with $\Sigma$. It should be noted that this value in general does not coincide with the value $\alpha^*$ obtained with the standard method by minimizing the Frobenius norm \cite{Schafer}.

\begin{figure}[ptb]
\begin{center}
              \includegraphics[scale=0.27]{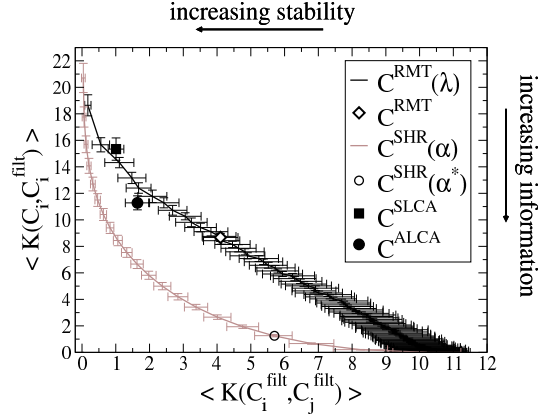}
              \caption{Stability of the filtered matrix ($x$ axis) against the amount of information about the correlation matrix that is retained in the filtered matrix ($y$ axis) for $n=100$ stocks of the NYSE in the period 2001-2003 ($T=748$). 
  The points labeled with ${\bf C}^{RMT}(\lambda)$ correspond to the filtering procedure keeping a fixed number of eigenvalues. The number of kept eigenvalues increases when one goes from the top left to the bottom right corner. The points labeled with ${\bf C}^{SHR}(\alpha)$ correspond to the shrinkage procedure (see Eq.~\ref{shrinkage}) and the parameter $\alpha$ goes from $0$ to $1$ when one goes from the bottom right to the top left corner.} 
              \label{NYSE}
\end{center}
\end{figure}

We now consider an application to a real system. We investigate the daily returns of $n=100$ highly capitalized stocks traded at the NYSE in the period 2001-2003 ($T=748$).  In Fig. ~\ref{NYSE} we show  the performance of different filtering procedures in the plane stability-information. First of all it is worth noting that no kink is observed when one varies the number of eigenvalues retained. This indicates that the block diagonal matrix is far from being a faithful representation of financial correlation matrices. RMT, SLCA and ALCA have different properties in terms of stability and information \cite{Tumminello2007}. SLCA is the most stable even if it is the least informative, whereas RMT is the least stable but the most informative. ALCA has intermediate properties both with respect to stability and to information.
As for the models the shrinkage seems to outperform the other filtering techniques, even if in this case a quantitative prediction of the optimal value of $\alpha$ is more difficult due to the non-Gaussianity of financial returns. This point will be discussed in the next section. 

%Finally the shrinkage filtering procedure is able to achieve a very good compromise between stability and information and the value $\alpha\simeq 0.4$ seems to be the optimal choice.   It is also interesting to note that there exist analytical methods to obtain the optimal value of $\alpha$ according to some cost function \cite{Schafer}. By using the standard quadratic (or Frobenius) norm we found that the optimal parameter is $\alpha^*=0.14$. This value, also shown in Fig.~\ref{NYSE}, is far from being optimal in the information-stability plane based on KL distance. 

\section{A first extension to non-Gaussian variables}

The results obtained so far are valid for multivariate Gaussian variables. However in many real systems the random variables of interest are non-Gaussian, and have often the property that the tails of the distribution are significantly fatter than in the Gaussian case. A paradigmatic example is financial price return discussed above. In this section we present some numerical results obtained for a specific class of non-Gaussian variables. A non-Gaussian multivariate distribution useful in describing financial returns is the multivariate Student's t-distribution \cite{Bouchaud}. 

The multivariate distribution is
\begin{equation}
P(x_1,x_2,...,x_n)=\frac{\Gamma(\frac{n+\mu}{2})}{\Gamma(\mu/2)\sqrt{(\mu \pi)^n | {\bf \Sigma}|}}\frac{1}{\left(1+\frac{1}{\mu}\sum_{i,j} x_i( {\bf \Sigma}^{-1})_{ij}x_j\right)^{\frac{n+\mu}{2}}}.
\label{studenteq}
\end{equation}
The parameter $\mu$ describes the tail behavior of the marginal distribution of any $x_i$ since $P(x_i)\sim x_i^{-1-\mu}$. A process distributed as Eq.~\ref{studenteq} can be obtained by setting  $x_i(t)=\sigma(t) \eta_i(t)$, where the $\eta$s are multivariate Gaussian variables with correlation matrix ${\bf \Sigma}$ and $\sigma(t)$ is a suitably distributed random variable. 

\begin{figure}[ptb]
\begin{center}
              \includegraphics[scale=0.27]{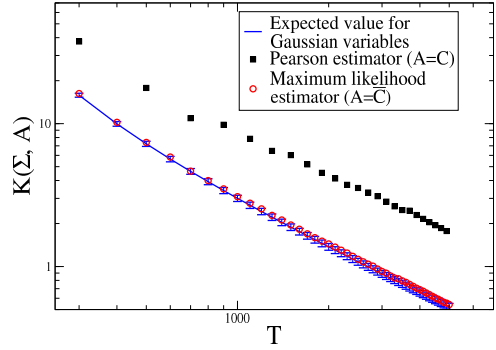}
              \caption{KL distance of Eq.~\ref{kullbackGaussian} between the model correlation matrix ${\bf \Sigma}$ and two estimators of the sample correlation matrix, specifically the Pearson estimator ${\bf C}$ and the maximum likelihood estimator ${\bf \bar C}$ of Eq.~\ref{mle}. The data are generated according to the multivariate Student's t-distribution of Eq.~\ref{studenteq} with $\mu=4$. The correlation matrix of the model is the one of a hierarchically nested model of 100 elements following the HNFM with 23 factors obtained in Ref. \cite{epl}. 
        } \label{student}
\end{center}
\end{figure}

In order to check whether the results on the KL distance for Gaussian distributions described above also hold for Student's t-distributions we have generated samples of $T$ records of a multivariate Student's t-distribution of $n=100$ variables. The correlation matrix of the underlying Gaussian variables is the HNFM described in Ref. \cite{epl} and this model is the same as the one used in Ref. \cite{Tumminello2007} and  in the right panel of Fig.~\ref{RMT}. We made this choice in order to have a correlation matrix  with a non trivial structure. By using Eq.~\ref{kullbackGaussian} we then compute the KL distance between the model correlation matrix and a sample correlation matrix obtained with the Pearson estimator. We compare this value with the expected value of $K({\bf \Sigma},{\bf C}_1)$ of Eq.~\ref{kullexpecSigS1} and we find that these values are significantly different.Specifically, the value of Eq.~\ref{kullbackGaussian}, obtained by using the sample Pearson correlation matrix, is larger than the expected value given by Eq.~\ref{kullexpecSigS1} (see Fig.~\ref{student}). At first sight this seems to indicate that the results on the KL distance for Gaussian distributions cannot be applied to non-Gaussian variables. However it is known \cite{Bouchaud} that the Pearson estimator of the correlation matrix is not the maximum likelihood estimator when the variables are non-Gaussian. In the case of the Student's t-distribution of Eq.~\ref{studenteq} there exists a recursive equation for the maximum likelihood estimator ${\bf \bar C}$ which is \cite{Bouchaud}
\begin{equation}
\bar C_{ij}=\frac{n+\mu}{T}\sum_{t=1}^T\frac{x_i(t)x_j(t)}{\mu+\sum_{pq} x_p(t)({\bf \bar C}^{-1})_{pq}x_q(t)}.
\label{mle}
\end{equation}
Fig. \ref{student} compares the KL distance of Eq.~\ref{kullbackGaussian} between the model correlation matrix ${\bf \Sigma}$ and the two estimators, specifically the Pearson estimator ${\bf C}$ and the maximum likelihood estimator ${\bf \bar C}$ of Eq.~\ref{mle}. The figure shows that, while $K({\bf \Sigma}, {\bf C})$ is not described by Eq.~\ref{kullbackGaussian}, the KL distance $K({\bf \Sigma}, {\bf \bar C})$ using the maximum likelihood estimator ${\bf \bar C}$ is well described  by Eq.~\ref{kullbackGaussian}. This result suggests that in some cases one can extend the results obtained for Gaussian variables to non-Gaussian variables provided that the maximum likelihood estimator and not the Pearson estimator is used in the computation of the KL distance. An analytical extension of the KL distance to non-Gaussian distributions is presented in Ref.~\cite{Biroli} of this issue. One of the obtained results confirms the conclusion drawn in this section about the Maximum Likelihood Estimator of Student correlation matrices.

\section{Conclusions}

We have considered the application of KL distance to the measurement of the performance of correlation matrix filtering procedures in giving reliable and stable estimates of the underlying correlation matrix.  Our analysis suggests that the optimal number of eigenvalues to be retained in filtering correlation matrices by mean of spectral procedures is close to the number of eigenvalues indicated by RMT. Our investigation of models also indicates that spectral filtering procedures are slightly more efficient in filtering ``separable'' systems, like those described by block diagonal models, than hierarchical clustering filtering procedures, whereas the latter work better for systems with a clear hierarchical structure of correlations. 
We have also shown that the shrinkage approach is very efficient in filtering a sample correlation matrix, although the estimate of the optimal shrinkage intensity in terms of the Frobenius norm is far from being optimal in terms of the KL distance. Finally, we have suggested a possible extension of our method to non-Gaussian variables.

\bigskip
{\bf Acknowledgments} We thank Jean-Philippe Bouchaud for very useful discussions and for sending us Ref.~\cite{Biroli} before publication. We acknowledge partial support from MIUR research project ``Dinamica di altissima frequenza nei mercati finanziari'' and NEST-DYSONET 12911 EU project.

%%%%%%%%%%%%%%%%%%%%%%%%%%%%%%%%%%%%%%%%%%%%%%%%%%%%%%%%%%%%%%%%%%%%%%%%%%%%%%%%%%%%%%%%%%%%%%%%%%%%%%%%%%
%%%%%%%%%%%%%%%%%%%%%%%%%%%%%%%%%%%%%%%%%%%%%%%%%%%%%%%%%%%%%%%%%%%%%%%%%%%%%%%%%%%%%%%%%%%%%%%%%%%%%%%%%%


\begin{thebibliography}{99}

\bibitem{brown} 
O. Alter, P. O. Brown and D. Botstein,
{\it Proc. Nat. Acad. Sci. USA} {\bf 97}, 10101 (2000). 

\bibitem{Eguiluz}V.M. Eguiluz, {\it et al.}
{\it Phys. Rev. Lett.} {\bf 94}, 018102 (2005).

\bibitem{Laloux1999}
L. Laloux, P. Cizeau, J.-P. Bouchaud, and M. Potters,
{\it Phys. Rev. Lett.} {\bf 83}, 1467-1470 (1999).

\bibitem{Plerou1999}
V. Plerou, P. Gopikrishnan, B. Rosenow, L. A. N. Amaral, and H. E. Stanley, 
{\it Phys. Rev. Lett.} {\bf 83}, 1471-1474 (1999).

\bibitem{Mantegna1999}
R. N. Mantegna, 
{\it Eur. Phys. J. B} {\bf 11}, 193-197 (1999).

\bibitem{Tumminello2007} M. Tumminello, F. Lillo, and R.N. Mantegna,  {\it Phys. Rev. E} {\bf 76} 031123 (2007).

\bibitem{Kullback}
S. Kullback and R. A. Leibler, 
{\it Ann. Math. Statist.} {\bf 22}, 79-86 (1951).

\bibitem{Cover}
T. M. Cover and J. A. Thomas, in
\emph{Elements of Information Theory}
(Wiley Interscience, New York, 1991).

\bibitem{Mardia}
K. V. Mardia, J. T. Kent, and J. M. Bibby in
{\it Multivariate Analysis}, 
(Academic Press, San Diego, CA, 1979).

\bibitem{Bartlett54}
M. S. Bartlett,
{\it J. Roy. Statist. Soc. B}, \textbf{16}, 296-298 (1954).

\bibitem{Metha90} M.L. Metha, 
{\it Random Matrices}
(Academic Press, New York, 1990).

\bibitem{Potters2005}
 M. Potters, J.-P. Bouchaud and L. Laloux, 
{\it Acta Phys. Pol. B}  \textbf{36} (9), 2767-2784 (2005). 


\bibitem{Anderberg}
M. R. Anderberg, in
\emph{Cluster Analysis for Applications} 
(Academic Press, New York, 1973).

\bibitem{epl}
M. Tumminello, F. Lillo and R. N. Mantegna, 
{\it Europhys. Lett.} {\bf 78}, 30006 (2007).

\bibitem{haff} L.R. Haff, {\it Ann. Statist.} {\bf 8} 586 (1980).

\bibitem{Ledoit03} 
O. Ledoit and M. Wolf,
{\it J. Mult. Analysis} {\bf 88}, 365 (2004).

\bibitem{Schafer}
J. Sch\"afer and K. Stimmer,
{\it Stat. Appl. Gen. Mol. Biol.} {\bf 4} (2005).

\bibitem{Bouchaud}
J.-P. Bouchaud and M. Potters,
{\it Theory of Financial Risk and Derivative Pricing},
(Cambridge University Press, 2003). 

\bibitem{Biroli} G. Biroli, J.-P. Bouchaud, and M. Potters,
{\it The Student ensemble of correlation matrices: eigenvalue spectrum and Kullback-Leibler entropy}, this volume.

\end{thebibliography}
\end{document}